\documentclass[conference]{IEEEtran}
\IEEEoverridecommandlockouts
\usepackage[pdftex]{graphicx}

\usepackage{cite}
\usepackage{amsmath,amssymb,amsfonts}
\usepackage{graphicx}
\usepackage{textcomp}
\usepackage{url}
\usepackage{listings}
\def\BibTeX{{\rm B\kern-.05em{\sc i\kern-.025em b}\kern-.08em
		T\kern-.1667em\lower.7ex\hbox{E}\kern-.125emX}}
\begin{document}

\newcommand*{\FFXVLongWithAbb}{FINAL FANTASY XV (FFXV)}
\newcommand*{\FFXVLongWithCite}{FINAL FANTASY XV (FFXV) \cite{FFXV}}
\newcommand*{\FFXV}{FFXV}
\newcommand*{\FFXVDevTeam}{the FINAL FANTASY XV development team}	

\lstset{%
	basicstyle={\footnotesize\ttfamily},%
	identifierstyle={\footnotesize},%
	commentstyle={\footnotesize\itshape},%
	keywordstyle={\footnotesize\bfseries},%
	ndkeywordstyle={\footnotesize},%
	stringstyle={\footnotesize\ttfamily},
	frame={tb},
	breaklines=true,
	columns=[l]{fullflexible},%
	numbers=left,%
	xrightmargin=0mm,%
	xleftmargin=5mm,%
	numberstyle={\footnotesize},%
	stepnumber=1,
	numbersep=2mm,%
	lineskip=0ex%
}

\title{
	Formal Verification for Node-Based Visual Scripts Using Symbolic Model Checking
}

\author{\IEEEauthorblockN{Isamu HASEGAWA}
\IEEEauthorblockA{
\textit{SQUARE ENIX CO., LTD.}\\ 
haseisam@square-enix.com}
\and
\IEEEauthorblockN{Tomoyuki YOKOGAWA}
\IEEEauthorblockA{
\textit{Okayama Prefectural University}\\
t-yokoga@cse.oka-pu.ac.jp}
}

\maketitle

\begin{abstract}
Visual script languages with a node-based interface have commonly been used in the video game industry.
We examined the bug database obtained in the development of \FFXVLongWithAbb, and noticed that several types of bugs were caused by simple mis-descriptions of visual scripts and could therefore be mechanically detected.

We propose a method for the automatic verification of visual scripts in order to improve productivity of video game development.
Our method can automatically detect those bugs by using symbolic model checking.
We show a translation algorithm which can automatically convert a visual script to an input model for NuSMV that is an implementation of symbolic model checking.

For a preliminary evaluation, we applied our method to visual scripts used in the production for \FFXV.
The evaluation results demonstrate that our method can detect bugs of scripts and works well in a reasonable time.
\end{abstract}


\section{Introduction}
\subsection{Background}\label{Sec:Background}
In the recent video game industry, game designers write game logic using script languages.
Although most of game designers are not familiar with writing programs, the use of visual script languages allow such designers to describe game logic, and thus can improve the productivity of game development.
In particular, visual script languages with a node-based interface are widely used in game development.

However, it is hard to maintain game logic written in visual script languages because such scripts can quickly become large and complicated during the development process, and thus become hard to be verified or modified, and very prone to human error. 

We examined the bug database obtained in the development of \FFXVLongWithCite, and noticed that several types of bugs were caused indeed by simple mis-descriptions of visual scripts.
A system that can automatically detect such mis-descriptions would had been a great help to our production.

Most visual script implementations could be treated as a kind of state machine \cite{HAREL1987}.
While model checking is a well-researched technique to automatically verify finite state machines \cite{Chan1998}\cite{Zhao2006}.
We thus propose in this paper a method for automatic verification of visual script notations with symbolic model checking \cite{Burch1992} for efficient game production.
Our main contributions are the following.
(1) To apply symbolic model checking to verify visual scripts, we provide a translation algorithm from a visual script description to an input model for NuSMV \cite{Cimatti1999}, that is an implementation of symbolic model checking.
(2) We show a preliminary evaluation of our method by applying it to visual scripts which are produced in the development of \FFXV, and demonstrate that most of the verification tasks are completed in a realistic amount of time.

This paper is an extended version of our work published at ICFEM 2019 \cite{Hasegawa2019}.
The rest of this paper is organized as follows.
Section \ref{Sec:Approach} explains the proposed method.
Section \ref{Sec:TranslationAlgorithm} provides the translation algorithm from a visual script to an input model which can be accepted to NuSMV.
We additionally provide the method for optimizing the SMV program in order to reduce the state space in Section \ref{Sec:StateSpaceReduction}.
We show the results of our preliminary evaluation in Section \ref{Sec:PreliminaryEvaluation} and conclude our work in Section \ref{Sec:Summary}.

\subsection{Model Checking}
Model checking is an automatic technique for verifying correctness properties of a finite-state system \cite{Clarke1999}.
The verification procedure is performed by an exhaustive search over the state space.
Since the size of the state space exponentially increases with the number of system components, it is difficult to apply model checking to large-scale systems.
Symbolic model checking can efficiently handle large-scale systems by replacing explicit state representation with boolean formula.

NuSMV \cite{Cimatti1999} is one of the most successful implementations of symbolic model checking.
The model verified by NuSMV is written by a specific input language (called SMV language).
The properties to be checked is expressed by temporal logic LTL (Linear Temporal Logic) \cite{Pnueli1981} and CTL (Computational Tree Logic) \cite{Ben-Ari1983}.

\begin{figure}[htb]
\begin{lstlisting}[]
MODULE main
VAR
  sw : {on, off};
ASSIGN
  init(sw) := {on, off};
  next(sw) := case
    sw = on : off;
    TRUE : sw;
  esac;
CTLSPEC AG (AF sw = on)
\end{lstlisting}
\caption{An example model described in SMV language}\label{Fig:SMVexample}
\end{figure}

Fig. \ref{Fig:SMVexample} is an example of an input model to NuSMV.
The input model described by SMV language is composed of variable declaration part (described by \texttt{VAR}) and transition relation definition part (described by \texttt{ASSIGN}).
The property is expressed as a LTL formula (described by \texttt{LTLSPEC}) or a CTL formula (described by \texttt{CTLSPEC}).

This example has one variable \texttt{sw} which may have one of the two values \texttt{on} and \texttt{off}.
In its initial state, either \texttt{on} or \texttt{off} is assigned to \texttt{sw} non-deterministically.
In the case that \texttt{sw} is \texttt{on}, \texttt{sw} becomes \texttt{off} in the next state, or \texttt{sw} does not change its value.
Thus the sequence of the value of \texttt{sw} can be either \texttt{on}, \texttt{off}, \texttt{off} $\ldots$ (when the initial value is \texttt{on}) or \texttt{off}, \texttt{off} $\ldots$ (when the initial value is \texttt{off}).
The CTL formula in this example has two CTL operators \textbf{AG} and \textbf{AF}.
\textbf{AG} represents ``in Any path'' and ``Globally,'' and \textbf{AF} represents ``in Any path'' and ``in the Future.''
This formula expresses the following property: the system always satisfies that \texttt{sw} necessarily becomes \texttt{on}.
When the model is inputted to NuSMV, NuSMV returns \texttt{FALSE} for this property because there is a path where \texttt{sw} continues to be \texttt{off}.
Fig. \ref{Fig:Counterexample} shows the result and the counterexample generated by NuSMV.
The counterexample shows the path where \texttt{sw} continues to be \texttt{off}.

\begin{figure}[htb]
\begin{lstlisting}[]
-- specification AG (AF sw = on)  is false
-- as demonstrated by the following
execution sequence
Trace Description: CTL Counterexample
Trace Type: Counterexample
-> State: 1.1 <-
sw = on
-- Loop starts here
-> State: 1.2 <-
sw = off
-> State: 1.3 <-
\end{lstlisting}
\caption{A counterexample generated by NuSMV}\label{Fig:Counterexample}
\end{figure}

\subsection{Related Work}
Video games essentially have a large number of combinations of internal states and external stimuli.
This makes it difficult to detect problems which come out under specific conditions by testing.
Model checking has been applied to video game developments since it can solve such problems by exhaustive verification.
Moreno-Ger et al. \cite{Moreno-Ger2009} proposed a method for verifying game scripts created in $\langle$e-Adventure$\rangle$ platform using NuSMV.
Radomski et al. \cite{Radomski2015} showed a framework in which video game logics are modeled by State Chart XML (SCXML) formalism and their properties can be checked by the SPIN model checker.
Rezin et al. \cite{Rezin2018} developed a method to model a multi-player game design as a Kripke structure and to verify it by NuSMV.
These studies show that applying model checking to video game development is very promising and application to game logic described by node-based visual scripts is also expected.

There have been a number of studies that have applied model checking to verification of node-based state transition system designs.
Statecharts and its variants, such as UML state machine \cite{Rumbaugh2004} and RSML (Requirements State Machine Language) \cite{Leveson1994}, are one of the most popular notations for describing state transition systems in a node-based manner.
Chan et al. \cite{Chan1998} provided a translation from RSML notation to a model described by SMV language.
This translation procedure encodes components of the inputted RSML by SMV variables and expresses changes of the components as transition relation.
Zhao et al. \cite{Zhao2006} studied representation of Statecharts step-semantics as a Kripke structure, which is a graph-based state transition representation,
and carried out verification using SMV model-checker.
Jussila et al. \cite{Jussila_i.:model} presented a representation of a subset of UML state machines as Promela which is an input language of SPIN model-checker.

In the semantics of Statecharts and its variants, their nodes represent states and only simple actions (enter/exit or do action in the case of UML state machine) can be assigned to each node.
While in the visual script notations that we focus, each node expresses some game logic computation which can be performed individually and can have a particular semantics.
Thus it is difficult to directly apply the existing procedures to the verification of such a visual script notation.
In this paper, we propose a method to translate from visual scripts to models by SMV language.

\section{Approach}\label{Sec:Approach}
\subsection{Motivating Example}\label{SubSec:TypicalBugs}

Many game development environments have their own visual scripting system such as Blueprint in Unreal Engine 4\cite{UE4:Blueprints}, Lumberyard ScriptCanvas\cite{Lumberyard:ScriptCanvas}, and the VFX development tool for \FFXV\cite{Hasegawa:2016:VEF:2897839.2927427}.
Although there are slight differences among each visual scripting systems, their syntax and semantics are basically the same. 
In this paper, readers can assume Blueprint \cite{UE4:Blueprints} as the visual scripting system since its syntax and semantics are very similar to our in-house visual scripting system.

In the development with node-based visual script languages, logic is described as a {\em node graph} which is composed of {\em nodes} and {\em edges}.
Nodes express values, variables, arithmetic operators, or control statements which correspond to if/while-statements in text-based script languages.
Since the purpose of visual scripts is to control game components such as sound, visual effect, and so on, many nodes express invocations of APIs for those components.
For example, ``Play SE'' node notifies sound component of the game system to start playing sound effect, ``Fade Out'' node notifies screen effect component to start fade-out effect \footnote{``fade out'' is a gradual transition from the game screen to blank image, used in movies, games, etc.}.
Edges connect nodes through input and output {\em ports}, and express data and control flows.

\begin{figure}[t]
	\begin{center}
		\includegraphics[width=1.0\linewidth]{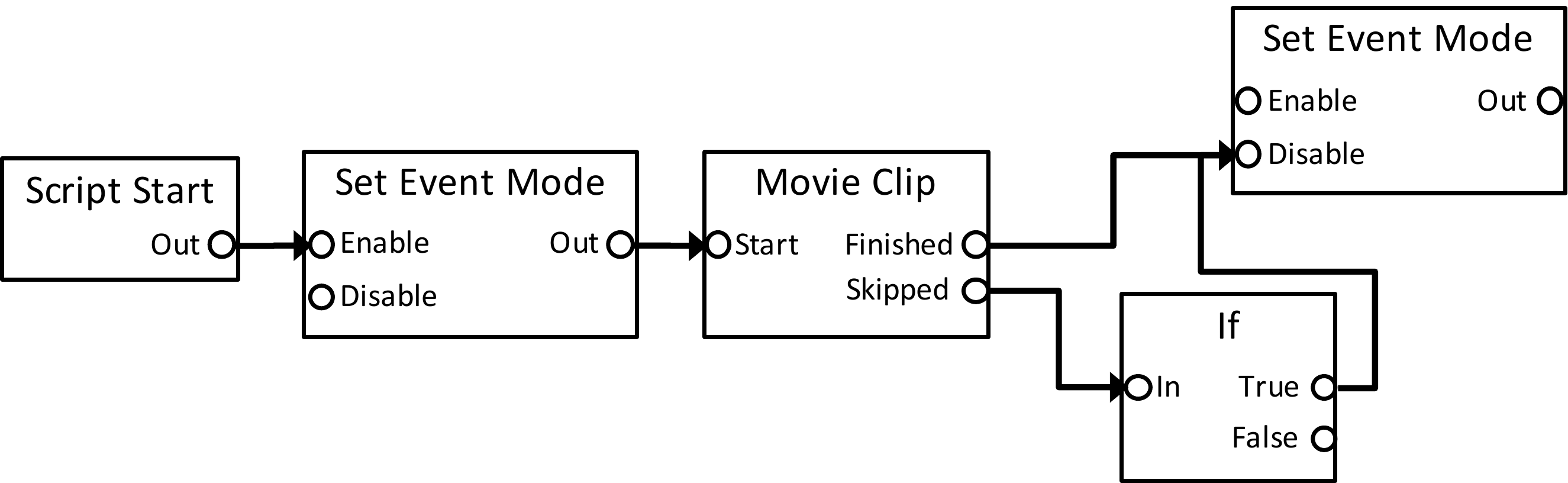}
	\end{center}
	\caption{An example of node-based visual script with a typical bug}\label{Fig:VisualScriptTypicalBug}
\end{figure}

Fig.\ref{Fig:VisualScriptTypicalBug} shows an example of visual script.
For simplicity, we omitted data flow edges such as the condition value inputted to \textsf{If} node.
This is because our method does not address the detection of bugs caused by an illegal data flow.

This example has the following behavior:
\begin{itemize}
	\item The \textsf{Set Event Mode} node changes the value of the global flag variable \textsf{event mode}.
	When it receives the input signal through the \textsf{Enable} or \textsf{Disable} port, the event mode flag becomes true or false respectively.
	This example includes two \textsf{Set Event Mode} nodes, and both of them modify the same variable instance because the \textsf{event mode} is a global variable referred from the whole game system.
	\item When the \textsf{Movie Clip} node receives an input signal through the \textsf{Start} port,
	it starts playing the movie clip,
	and sends output signal through the \textsf{Finished} port when the clip finishes.
	If the movie clip is skipped by a player, it sends output signal through the \textsf{Skipped} port instead of the \textsf{Finished} port.
	\item The \textsf{If} node performs conditional branching.
	Depending on the condition value, it outputs signal through the \textsf{True} or \textsf{False} port.
	Its condition value is inputted through data flow port (As stated above, we omit such ports).	
	\item The value of \textsf{event mode} must be true while the movie clip is playing in order to change the game state appropriately.
	(For instance, the gamepad is disabled during the movie.)
	On the other hand, it must be false when the movie clip is not playing.
\end{itemize}

Note that the \textsf{Movie Clip} node has internal states and sends \textsf{Finished} and \textsf{Skipped} signal depending on its internal state.
Since its output signals are activated independently of the original control flow through the input port, there can be multiple activated nodes and multiple activated control signals simultaneously in the graph.
This is one of the major differences between visual script languages and Statecharts, and the reason why prior researches cannot be applied directly to the verification of visual scripting languages.

This example contains a bug that actually occurred frequently during the development of \FFXV.
As you can see, the \textsf{False} port of the \textsf{If} node is not connected to any port.
Therefore, in the case that the \textsf{If} node branches to \textsf{False} after the \textsf{Movie Clip} node outputs \textsf{Skipped} signal, the \textsf{event mode} flag will not be changed and will remain true.
As a result, the \textsf{event mode} flag will remain true even though the movie clip has finished playing, resulting in incorrect behavior.

There were a wide variety of similar bugs during the development of \FFXV, e.g. ``BGM is not changed correctly in some cases,'' ``Enemy characters never respawn in a specific condition,'' and so on.
Although these bugs are caused by trivial mis-descriptions such as missing one node or edge, it is tough to find those bugs by visual inspection.
This is due to the fact that many game logic scripts are described by game designers who are not familiar with programming, and thus the script becomes large and complicated.
Our goal is to detect those large amounts of trivial but hard-to-find bugs automatically and exhaustively.
Since our products already have a lot of massive scripts, we should cover not only newly written scripts but also those existing scripts.

\subsection{Overview}
\begin{figure}[h]
	\begin{center}
		\includegraphics[width=1.0\linewidth]{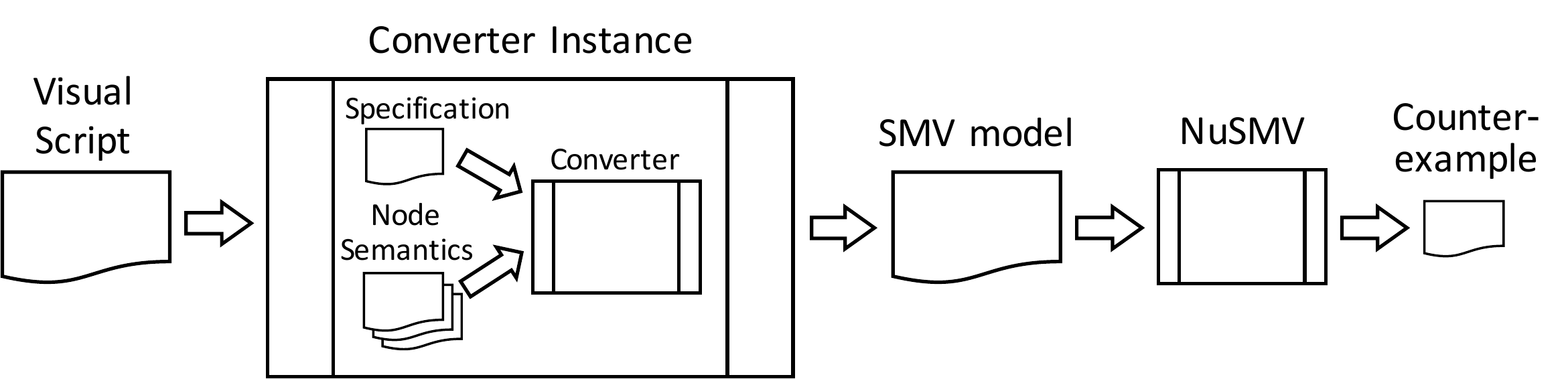}
	\end{center}
	\caption{System overview}\label{Fig:SystemOverview}
\end{figure}

Fig. \ref{Fig:SystemOverview} shows the system overview of the visual script verification environment with NuSMV.
This environment carries out verification by converting a visual script into an SMV model.
First, the system generates a converter instance from specifications to be checked and the corresponding node semantics.
Then the visual script is converted into an SMV model by using the converter instance.
NuSMV can verify whether the inputted visual script satisfies the specifications or not.
When the specifications are not satisfied, NuSMV outputs counterexamples.

In our system, specifications and node semantics need to be written manually.
However, once they are given, the verification process can be fully automated.
Specifications for detecting typical bugs can be represented by simple formulas in temporal logic.
Also, since the behavior of many nodes is common, many nodes can be converted to SMV models by writing a single node semantics.

\subsubsection{Specification}\label{Sec:Specification}
In our visual script verification environment, specifications are expressed by propositional temporal formulas.
Since NuSMV supports temporal formulas written in CTL and LTL, the specification must be written in those logics in our system as well.

For example, in order to detect the bugs explained in Fig. \ref{SubSec:TypicalBugs}, we provide the specification expressed as the following CTL formula:
\[ AG (\textsf{event mode} = true \rightarrow AF (\textsf{event mode} = false)). \]
This formula represents that it is always satisfied that \textsf{event mode} flag must eventually become false if the flag is true.

\subsubsection{Node Semantics}\label{SubSec:NodeSemantics}
Each node has input or output ports, and internal states (if the node has state transition).
It also can change the values of global variables.
The output signal of the node depends on its input signal and its internal state.
The internal state of the node changes according to its input signals and its current state.
Thus the transition relation of the value of the output signal and the internal state of the node is generally expressed the following formula:
\begin{align*}
out'   & \equiv f_{out}(in, state), \\
state' & \equiv f_{state}(in, state).
\end{align*}
$in$, $out$ and $state$ denotes the value of the input signal, output signal and internal state, respectively.
The primed symbol denotes the value after the transition.

Also, the transition relation of the values of global variable changed by the node is similarly expressed the following formula:
\begin{align*}
var'   & \equiv f_{var}(in, state, var).
\end{align*}
$var$ denotes the values of global variable.
For example, the \textsf{Set Event Mode} node has the following relation:
\begin{align*}
out'   & \equiv \text{if } (in = \textsf{Enable} \mid in = \textsf{Disable}) \text{ then } \textsf{Out}\\
& \quad \text{else } none, \\
\textsf{event mode}'   & \equiv \text{if } (in = \textsf{Enable}) \text{ then } true\\
& \quad \text{else if } (in = \textsf{Disable}) \text{ then } false \\
& \quad \text{else } \textsf{event mode}.
\end{align*}
We refer the transition relation of the elements involved in the node as \emph{node semantics}.
Initial values of elements are also defined in node semantics.

In our system, node semantics are given as templates for an SMV program that define the transition relation.
Fig. \ref{Fig:NodeSemanticsExample} shows the example of node semantics for the \textsf{Set Event Mode} node.
Note that this definition only depends on the elements involved in the node.
We thus can describe node semantics independent from graph structure.

Our translation algorithm generates an SMV program by appropriately assigning variables to the node semantics of all nodes.
For example, since Fig. \ref{Fig:VisualScriptTypicalBug} has two \textsf{Set Event Mode} nodes, our algorithm applies the node semantics to these nodes and assigns different variables to them.

\begin{figure}[htb]
\begin{lstlisting}[basicstyle=\sffamily]
@SetEventMode:define:output_variable
  init(<output_variable>) := none;
  next(<output_variable>) := case
    <input_variable> = Enable | <input_variable> = Disable : Out;
    TRUE : none;
  esac;
@SetEventMode:rule:EventMode
  <input_variable> = Enable : true;
  <input_variable> = Disable : false;
\end{lstlisting}
\caption{An example of node semantics}\label{Fig:NodeSemanticsExample}
\end{figure}

\subsection{Translating Example}\label{Sec:ModelOverview}
We first show the overview of the translation using the example visual script shown in Fig. \ref{Fig:VisualScriptTypicalBug}.
Fig. \ref{Fig:ModelExample} is an SMV model obtained from the visual script.

\newcounter{exmodelcounter}
\setcounter{exmodelcounter}{0}
\begin{figure*}[!htb]
\begin{lstlisting}[basicstyle=\ttfamily\scriptsize, frame=single, escapechar=@, columns=fixed, basewidth=0.65em]
MODULE main
VAR
  ScriptStart1Out  : {none, Out};
  SetEventMode2In  : {none, Enable, Disable};                     -- (@\refstepcounter{exmodelcounter}{\label{Model:DeclInOutVar}}\theexmodelcounter@)
  SetEventMode2Out : {none, Out};                                 -- (@\theexmodelcounter@)
  MovieClip3In     : {none, Start};
  MovieClip3Out    : {none, Finished, Skipped};
  MovieClip3State  : {Stopped, Playing, Finished, Skipped};       -- (@\refstepcounter{exmodelcounter}{\label{Model:DeclStateVar}}\theexmodelcounter@)
  SetEventMode4In  : {none, Enable, Disable};
  SetEventMode4Out : {none, Out};
  If5In            : {none, In};
  If5Out           : {none, True, False};
  EventMode        : {true, false};                               -- (@\refstepcounter{exmodelcounter}{\label{Model:DeclScriptVar}}\theexmodelcounter@)
FAIRNESS MovieClip3State = Stopped;                               -- (@\refstepcounter{exmodelcounter}{\label{Model:FairnessCond}}\theexmodelcounter@)
ASSIGN
  init(ScriptStart1Out) := Out;                                   -- (@\refstepcounter{exmodelcounter}{\label{Model:DefEntryPoint}}\theexmodelcounter@)
  next(ScriptStart1Out) := none;                                  -- (@\theexmodelcounter@)
  init(SetEventMode2In) := none;                                  -- (@\refstepcounter{exmodelcounter}{\label{Model:DefInVarInit}}\theexmodelcounter@)
  next(SetEventMode2In) := case
    ScriptStart1Out = Out : Enable;                               -- (@\refstepcounter{exmodelcounter}{\label{Model:DefInVar}}\theexmodelcounter@)
    TRUE                  : none;
  esac;
  init(SetEventMode2Out) := none;                                 -- (@\refstepcounter{exmodelcounter}{\label{Model:DefOutVarInit}}\theexmodelcounter@)
  next(SetEventMode2Out) := case                                  -- (@\refstepcounter{exmodelcounter}{\label{Model:DefOutVar}}\theexmodelcounter@)
    SetEventMode2In = Enable | SetEventMode2In = Disable : Out;   -- (@\theexmodelcounter@)
    TRUE                                                 : none;  -- (@\theexmodelcounter@)
  esac;
  init(MovieClip3In) := none;
  next(MovieClip3In) := case
    SetEventMode2Out = Out : Start;
    TRUE                   : none;
  esac;
  init(MovieClip3Out) := none;
  next(MovieClip3Out) := case
    MovieClip3State = Finished : Finished;                        -- (@\refstepcounter{exmodelcounter}{\label{Model:DefStateOutVar}}\theexmodelcounter@)
    MovieClip3State = Skipped  : Skipped;                         -- (@\theexmodelcounter@)
    TRUE                       : none;
  esac;
  init(MovieClip3State) := Stopped;                               -- (@\refstepcounter{exmodelcounter}{\label{Model:DefStateVarInit}}\theexmodelcounter@)
  next(MovieClip3State) := case
    MovieClip3In = Start      : Playing;                          -- (@\refstepcounter{exmodelcounter}{\label{Model:DefStateVarPlaying}}\theexmodelcounter@)
    MovieClip3State = Playing : {Playing, Finished, Skipped};     -- (@\refstepcounter{exmodelcounter}{\label{Model:DefStateVarFinished}}\theexmodelcounter@)
    TRUE                      : Stopped;                          -- (@\refstepcounter{exmodelcounter}{\label{Model:DefStateVarToInit}}\theexmodelcounter@)
  esac;
  init(SetEventMode4In) := none;
  next(SetEventMode4In) := case
    MovieClip3Out = Finished : Disable;
    If5Out = True            : Disable;
    TRUE                     : none;
  esac;
  init(SetEventMode4Out) := none;
  next(SetEventMode4Out) := case                                  -- (@\refstepcounter{exmodelcounter}{\label{Model:DefOutVar2}}\theexmodelcounter@)
    SetEventMode4In = Enable | SetEventMode4In = Disable : Out;   -- (@\theexmodelcounter@)
    TRUE                                                 : none;  -- (@\theexmodelcounter@)
  esac;
  init(If5In) := none;
  next(If5In) := case
    MovieClip3Out = Skipped : In;
    TRUE                    : none;
  esac;
  init(If5Out) := none;
  next(If5Out) := case
    If5In = In : {True, False};                                   -- (@\refstepcounter{exmodelcounter}{\label{Model:DefOutVarBranch}}\theexmodelcounter@)
    TRUE       : none;
  esac;
  init(EventMode) := false;
  next(EventMode) := case                                         -- (@\refstepcounter{exmodelcounter}{\label{Model:DefScriptVar}}\theexmodelcounter@)
    SetEventMode2In = Enable | SetEventMode4In = Enable   : true; -- (@\theexmodelcounter@)
    SetEventMode2In = Disable | SetEventMode4In = Disable : false;-- (@\theexmodelcounter@)
    TRUE                      : EventMode;                        -- (@\theexmodelcounter@)
  esac;

CTLSPEC AG(EventMode = true -> AF(EventMode = false))             -- (@\refstepcounter{exmodelcounter}{\label{Model:Spec}}\theexmodelcounter@)
\end{lstlisting}
	\caption{Converted SMV model}\label{Fig:ModelExample}
\end{figure*}

\subsubsection{Variables}
We prepare four types of SMV variables to describe the behavior of a visual script.
\begin{itemize}
	\item {\em Input} and {\em output variables} represent activated ports of each node.
	Since only one input and output port of the nodes can be activated at the same time, we declare one input and output variable for each node.
	For example, the input and output variable for the leftmost \textsf{Set Event Mode} node is defined as \texttt{SetEventMode2In} and \texttt{SetEventMode2Out}, respectively (see Fig. \ref{Fig:ModelExample} (\ref{Model:DeclInOutVar})).
	The domain of an input and output variable is defined as the names of its input and output port and special value \texttt{none} which denotes that no port is activated.
	Since \textsf{Set Event Mode} node has two input ports \textsf{Enable} and \textsf{Disable}, the input variable \texttt{SetEventMode2In} has the domain \texttt{\{none, Enable, Disable\}}.
	The initial value of the input and output variable is set to \texttt{none} (see Fig. \ref{Fig:ModelExample} (\ref{Model:DefInVarInit}) and (\ref{Model:DefOutVarInit})).
	
	\item {\em Script variables} represent global variables of the visual script and information of external components that the visual script interacts with.
	Specifications are often expressed as correct behaviors of those variables.
	For example, the script variable for the global variable \textsf{event mode}, which is a flag variable of the external game system that the visual scripts interact with, is defined as \texttt{EventMode} (see Fig. \ref{Fig:ModelExample} (\ref{Model:DeclScriptVar})).
	
	\item {\em State variables} represent the internal state of each node which has state transition semantics.
	For example, the state variable for \textsf{Movie Clip} node is defined as \texttt{MovieClip3State} (see Fig. \ref{Fig:ModelExample} (\ref{Model:DeclStateVar})).
	The domain of a state variable is defined as a set of its internal states.
	Since \textsf{Movie Clip} node has four internal states, the state variable \texttt{MovieClip3State} has the domain \texttt{\{Stopped, Playing, Finished, Skipped\}}.
\end{itemize}

\subsubsection{Control Flows}
Control flows in a visual script are represented as propagation of signals through the edges between the nodes.
When an output port is activated, then the input port connected through the edge to the output port is activated.
Therefore, we can describe such propagation of signal in an SMV model as value assignments for input variables according to the values of output variables.

For example, the input port \textsf{Enable} of the leftmost \textsf{Set Event Mode} node is connected to the output port \textsf{Out} of \textsf{Script Start} node.
Thus the value of the input variable \texttt{SetEventMode2In} changes to \texttt{Enable} when the output variable \texttt{ScriptStart1Out} has the value \texttt{Out} (see Fig. \ref{Fig:ModelExample} (\ref{Model:DefInVar})).

The signal propagation representing the control flow of a visual script is reproduced as value transitions in an SMV model.
For example, the following is one of the control flows of the visual script in Fig. \ref{Fig:VisualScriptTypicalBug}:
\begin{align*}
\texttt{ScriptStart:Out} & \rightarrow \texttt{SetEventMode:Enable} \rightarrow \\
\texttt{SetEventMode:Out} & \rightarrow \texttt{MovieClip:Start} \rightarrow\\
\texttt{MovieClip:Skipped} & \rightarrow \texttt{If:In} \rightarrow \\
\texttt{If:False}& .
\end{align*}
This control flow is obtained as value transitions of SMV variables shown in Fig. \ref{Fig:NodeModel1Behavior}.

\begin{figure}[htb]
\begin{lstlisting}[]
-> State: 1.1 <-
ScriptStart1Out = Out
SetEventMode2In = none
...
MovieClip3State = Stopped
EventMode = false
-> State: 1.2 <-
ScriptStart1Out = none
SetEventMode2In = Enable
-> State: 1.3 <-
SetEventMode2In = none
SetEventMode2Out = Out
EventMode = true
-> State: 1.4 <-
SetEventMode2Out = none
MovieClip3In = Start
-> State: 1.5 <-
MovieClip3In = none
MovieClip3State = Playing
-> State: 1.6 <-
MovieClip3State = Skipped
-> State: 1.7 <-
MovieClip3Out = Skipped
MovieClip3State = Stopped
-> State: 1.8 <-
MovieClip3Out = none
If5In = In
-> State: 1.9 <-
If5In = none
If5Out = False
\end{lstlisting}
\caption{Value transitions of the example control flow obtained by NuSMV}\label{Fig:NodeModel1Behavior}
\end{figure}

\subsubsection{Nodes}
Each node activates its output ports and changes the value of global variables according to the input signals and its internal states.
Node semantics specifies such a behavior of a node and is described as value assignments for output, script and state variables in an SMV model.

\par
\noindent
\textbf{\textsf{Set Event Mode} node}\quad
The \textsf{Set Event Mode} changes the value of global variable to true or false when receiving input signal through \textsf{Enable} or \textsf{Disable} port, respectively.
Then it immediately sends the output signal through \textsf{Out} port.
Such behavior is described as the assignment of values for the output variables \texttt{SetEventMode2Out}, \texttt{SetEventMode4Out} and the script variable \texttt{EventMode}.
Both \texttt{SetEventMode2Out} and \texttt{SetEventMode4Out} change to \texttt{Out} when their input becomes \texttt{Enable} or \texttt{Disable} (see Fig. \ref{Fig:ModelExample} (\ref{Model:DefOutVar}) and (\ref{Model:DefOutVar2})). 
\texttt{EventMode} changes to \texttt{true} or \texttt{false} when the input of either of those nodes is \texttt{Enable} or \texttt{Disable}, respectively (see Fig. \ref{Fig:ModelExample} (\ref{Model:DefScriptVar})).

\par
\noindent
\textbf{\textsf{Movie Clip} node}\quad
Initially, the \textsf{Movie Clip} node does not play the movie clip and starts playing when it receives the input signal \textsf{Start}
Then if the game player skips playing the movie, the output signal \textsf{Skipped} is activated; if not, the output signal \textsf{Finished} is activated after the movie ends.
With this behavior in mind, the node has the following four internal states:
\begin{itemize}
	\item \textsf{Stopped}: the node is in the initial state and does not play the movie clip.
	\item \textsf{Playing}: the node is playing the movie clip.
	\item \textsf{Finished}: the movie clip has been finished, and the node sends the output signal \textsf{Finished}.
	\item \textsf{Skipped}: the movie clip is skipped, and the node sends the output signal \textsf{Skipped}.
\end{itemize}
Thus the node semantics of the \textsf{Movie Clip} node is defined as follows:
\begin{align*}
out'   & \equiv f_{out}(state)\\
& = \text{if } state = \textsf{Finished} \text{ then } \textsf{Finished}\\
& \quad \text{else if } state = \textsf{Skipped} \text{ then } \textsf{Skipped}\\
& \quad \text{else } none,\\
state' & \equiv f_{state}(in, state)\\
& = \text{if } in = \textsf{Start} \text{ then } \textsf{Playing}\\
& \quad \text{else if } state = \textsf{Playing}\\ 
& \quad \quad \text{then } \{\textsf{Playing}, \textsf{Finished}, \textsf{Skipped}\}\\
& \quad \text{else } \textsf{Stopped}
\end{align*}

We can describe the behavior of the node as the assignment of values of output variable \texttt{MovieClip3Out} and state variable \texttt{MovieClip3State} as follows:
\begin{itemize}
	\item The initial state is \textsf{Stopped} (see Fig. \ref{Fig:ModelExample} (\ref{Model:DefStateVarInit})).
	\item When receiving the input signal \textsf{Start}, the state changes to \textsf{Playing} (see Fig. \ref{Fig:ModelExample} (\ref{Model:DefStateVarPlaying})).
	\item When the state is \textsf{Playing}, it changes to either \textsf{Playing}, \textsf{Finished}, or \textsf{Skipped} non-deterministically (see Fig. \ref{Fig:ModelExample} (\ref{Model:DefStateVarFinished})).
	This non-deterministic choice represents the behavior of the game player, who may skip the movie or wait until it is finished playing.
	\item When the state changes to \textsf{Finished} or \textsf{Skipped}, the node sends the output signal \textsf{Finished} or \textsf{Skipped} respectively (see Fig. \ref{Fig:ModelExample} (\ref{Model:DefStateOutVar})).
	Then the state is back to \textsf{Stopped} (see Fig. \ref{Fig:ModelExample} (\ref{Model:DefStateVarToInit})).
\end{itemize}

\par
\noindent
\textbf{\textsf{If} node}\quad
The \textsf{If} node branches \textsf{True} or \textsf{False} according to the condition value.
Since we do not consider data flow and external behavior which affects the condition value, we model this branch is to decide the output signal non-deterministically.
Thus the value \textsf{True} or \textsf{False} is assigned to the output signal of the node non-deterministically (see Fig. \ref{Fig:ModelExample} (\ref{Model:DefOutVarBranch})).
NuSMV can check the both branch of \textsf{True} and \textsf{False} exhaustively.

\par
\noindent
\textbf{\textsf{Script Start} node}\quad
The \textsf{Script Start} node is an entry point node and sends the output signal once in the initial state.
Thus the initial value of the output signal of the node is set to \texttt{Out} and from then on, it will always be \texttt{none} (see Fig. \ref{Fig:ModelExample} (\ref{Model:DefEntryPoint})).

\subsubsection{Specification}\label{Sec:SpecificationExample}
Specifications are expressed as temporal formulas over script variables.
As stated in \ref{Sec:Specification}, the specification formula that can detect the expected bug is expressed as a CTL formula.
In an SMV model, such formula is annotated by \texttt{CTLSPEC} statement (see Fig. \ref{Fig:ModelExample} (\ref{Model:Spec})).

\subsubsection{Counterexample}\label{SubSec:CounterExample}
As we stated in \ref{Sec:ModelOverview}, a control flow of a node graph correspond to value transitions of input and output variables.
If the property given by \texttt{CTLSPEC} is violated, NuSMV generates a counterexample which indicates the witness of property violation.
Since the counterexample can be obtained as the form of the value transitions of SMV variables, we can identify the control flow which causes the violation from the counterexample.

For example, NuSMV can detect the bug and generates a counterexample that shows a value transition shown in Fig. \ref{Fig:NodeModel1Behavior}.
It means that the control flow through \textsf{Skipped} port of \textsf{Movie Clip} node and \textsf{False} port of \textsf{If} node causes violation of the specification.
Thus we can detect a bug stated in \ref{SubSec:TypicalBugs}.

\section{Translation Algorithm}\label{Sec:TranslationAlgorithm}
\subsection{Translation Overview}
An overview of the procedure for translating a visual script to an SMV program is shown below.
Given specifications and node semantics, this translation procedure can be fully automated.

\begin{enumerate}
	\item Declare SMV variables according to node semantics.
	\item Describe control flows as transitions of input variables according to the graph structure.
	\item Describe behaviors of nodes as transitions of output and state variables according to node semantics.
	\item Describe changes of script variables according to node semantics.
	\item Annotate specifications.
\end{enumerate}

\subsection{Variable Declaration}\label{SubSec:VariableDeclaration}
First we declare SMV variables according to given node semantics.
As stated above, the SMV model of a visual script has four types of SMV variables.\\

\par
\noindent
\textbf{input variables}\quad
For each node, its input signal is represented as an input variable that has as its domain the name of its input ports and \texttt{none} denoting an undefined value.
The initial value of an input variable is set to \texttt{none}.

\par
\noindent
\textbf{output variables}\quad
Similar to input variables, an output signal is represented as an output variable that has as its domain the name of its output ports and \texttt{none}.
The initial value of an output variable is also set to \texttt{none} (except those of entry point nodes).

\par
\noindent
\textbf{state variables}\quad
In the case of a node with state transition, its internal state is represented as a state variable.
The domain of the variable is exactly the set of internal states of the node.
The initial value of the state variable of such node is defined based on its node semantics.

\par
\noindent
\textbf{script variables}\quad
A global variable is represented as a script variable.
The domain of the variable is defined according to the node semantics that refer the change of the script variable.\\

As we stated in section \ref{SubSec:TypicalBugs}, multiple nodes in a visual script can work in parallel.
It means that a single node may also receive multiple input signals simultaneously.
However, the definition of input variables described above assumes that a node receives only one input signal at the same time, in which case the other input signals are ignored.
This might cause an incorrect behavior if some nodes are assumed to handle multiple input signals at the same time (fortunately these are very rare though).
To avoid this problem, for nodes whose semantics require processing multiple input signals simultaneously, we define one input variable for each input port.

\subsection{Translating Control Flow}\label{SubSec:InputVariables}
In node-style visual script, its control flow is represented as propagation of signals through edges of the graph.
Since signals are described as \emph{input} and \emph{output variables}, propagation of a signal between an input port and an output port is described as a change of the input variable based on the output variable.
Thus in the SMV model, a control flow through edges is described as definitions of transitions of input variables.

Assume that the node \textsf{NodeA} has two input ports \textsf{InA1} and \textsf{InA2}, each with edges to the output port \textsf{OutB} of the node \textsf{NodeB} and the output port \textsf{OutC} of the node \textsf{NodeC}.
The transition of the input variable \texttt{NodeAIn} for the node \textsf{NodeA} is defined by the following steps:
\begin{enumerate}
	\item Define the initial value of \texttt{NodeAIn} as \texttt{none}.
	\item For each edge between the input port \textsf{In} and the output port \textsf{Out}, add the transition ``If \textsf{Out} is activated then \textsf{In} will be activated in the next state'' to the \texttt{next} description of \texttt{NodeAIn}.
	\item Add the default transition rule that describes that any output ports are not activated.
\end{enumerate}
The SMV program for the input variable \texttt{NodeAIn} is obtained as follows.
\begin{lstlisting}[
numbers=none,
frame=none,
basicstyle=\ttfamily,
ndkeywordstyle=\ttfamily,
stringstyle=\ttfamily,
identifierstyle=\ttfamily,
xleftmargin=5mm,
lineskip=0mm,
]
ASSIGN
  init(NodeAIn) := none;
  next(NodeAIn) := case
    NodeBOut = OutB : InA1;
    NodeCOut = OutC : InA2;
    TRUE : NodeAIn;
  esac;
\end{lstlisting}
All input variables can be automatically defined according to the graph structure of the visual script.

\subsection{Specifying Node Semantics}\label{SubSec:SpecifyingNodeSemantics}
As we stated in section \ref{SubSec:NodeSemantics}, node semantics is defined as the transition relation of elements involved in the node.
We now classify the node semantics to several types and show how to define the transition relation according to the type of node.
Once the transition relation is obtained, it is easy to create templates for an SMV program from it.

Nodes are classified into those with a single output and those with multiple outputs.
An entry point node is a special kind of node that has a single output.
Nodes with multiple outputs are classified into those with conditional branches and those with internal states (or both).

\subsubsection{Node with Single Output}\label{SubSec:SemanticsOutputVariables}
A node with a single output port sends an output signal immediately upon receiving an input signal. 
Thus the node semantics of the node that has a single output port \textsf{Out} is defined as follows:
\begin{align*}
out'   & \equiv f_{out}(in)\\
& = \text{if } in \neq none \text{ then } \textsf{Out} \text{ else } none.
\end{align*}

An entry point node activates its output signal only in the initial state and not thereafter.
Thus the node semantics of the node is defined as follows:
\begin{align*}
out'   & \equiv none.
\end{align*}

\subsubsection{Node with Conditional Branches}\label{SubSection:ConditionalBranches}
Some nodes select output signals according to its conditional value.
Since we do not consider data flow that affects the conditional value, we model the output signal to be selected non-deterministically.
Thus the node semantics of the node that has $n$ output ports \textsf{Out1}, \textsf{Out2}, $\ldots$ and $\textsf{Out}n$ is defined as follows:
\begin{align*}
out'   & \equiv f_{out}(in)\\
& = \text{if } in \neq none \text{ then } \{\textsf{Out1}, \textsf{Out2}, \ldots \textsf{Out}n\}\\
& \quad \text{ else } none.
\end{align*}
The transition relation of the node with three or more output ports can be defined similarly.

\subsubsection{Node with State Transitions}\label{SubSection:StateTransition}
Some nodes have internal states and send output signals depending on its internal state.
The internal state transitions depending on the current state and the input signal.
In most nodes, its output signal depends only on its internal state.
Thus, regarding with the output signal, the node semantics of the node with state transitions has the following form:
\begin{align*}
out'   & \equiv f_{out}(state) \\
& = \text{if } state = state_1 \text{ then } out_1\\
& \quad \text{ else if } state = state_2 \text{ then } out_2\\
& \quad \cdots\\
& \quad \text{ else } none,
\end{align*}
where $out_i$ denotes an output signal of the node, and $state_i$ is the internal state in which the node sends the output signal $out_i$.
$out_i$ can be a set of output signals to specify a non-deterministic choice on sending output signals.

Besides, in many cases, the next internal state is associated to either its input signal or the current internal state.
Thus the node semantics for the internal state has the following form:
\begin{align*}
state' & \equiv f_{state}(in, state)\\
& = \text{if } in = in_1 \text{ then } state_1^i\\
& \quad \text{ else if } in = in_2 \text{ then } state_2^i\\
& \quad \cdots\\
& \quad \text{ else if } state = state_1 \text{ then } state_1^s\\
& \quad \text{ else if } state = state_2 \text{ then } state_2^s\\
& \quad \cdots\\
& \quad \text{ else } state_n.
\end{align*}
If the input signal is $in_j$, then the internal state transitions to $state_j^i$.
On the other hand, when the node has no input signal and its internal state is $state_k$, the internal state transitions to $state_k^s$.
Similar to $f_{out}$, $state_j^i$ and $state_k^s$ can be a set of internal states to express a nondeterministic choice on state transition.

In the case of nodes with nondeterministic state transitions, such as \textsf{Movie Clip} node, only certain state transitions may continue to be selected.
In the example of \textsf{Movie Clip} node, we allow a sequence in which its internal state continues to be \textsf{Playing}.
However, such sequences often need to be excluded from the sequences to be checked.
To avoid such a problem, we introduce a fairness constraint which restricts the scope of verification to only ``fair'' state transition sequences.
Since our model intends that all nodes with state transitions eventually return to its initial state, we mechanically add for each state variable the fairness constraint that claims the state variable infinitely often has its initial value.
By adding this constraint, the behavior where the node never returns to the initial state is not considered in verification by NuSMV.
Returning to the example of \textsf{Movie Clip} node, we add the constraint by using \texttt{FAIRNESS} description to guarantee that the state variable \texttt{MovieClip3State} infinitely often has its initial value \texttt{Stopped} (see Fig. \ref{Fig:ModelExample} (\ref{Model:FairnessCond})).

\subsubsection{Node with Custom Semantics}\label{SubSection:CustomSemantics}
Some nodes do not follow the form described above, and their output signals and internal states are determined according to semantics specific to the node.
We need to provide \emph{custom semantics} for such nodes.
In general, custom semantics can be described in the following form:
\begin{align*}
out'   & \equiv f_{out}(in, state) \\
& = \text{if } cond_1(in, state) \text{ then } out_1\\
& \quad \text{ else if } cond_2(in, state) \text{ then } out_2\\
& \quad \cdots\\
& \quad \text{ else } none,\\
state' & \equiv f_{state}(in, state)\\
& = \text{if } cond_1(in, state) \text{ then } state_1\\
& \quad \text{ else if } cond_2(in, state) \text{ then } state_2\\
& \quad \cdots\\
& \quad \text{ else } state_n,
\end{align*}
where $cond_i(in, state)$ is the condition over the input signal and the internal state.

\subsection{Representing Script Variables}\label{SubSec:ScriptVariables}
Script variables represent global variables used in visual scripts and states of external components that visual scripts interact.
By defining script variables and describing the conditions for those variables, we can verify those conditions with NuSMV.
Note that we need not to define all the variables in visual scripts, but minimum variables that we want to verify in the specification.

Since script variables can be modified by multiple nodes, it is necessary to integrate their transition relations defined in each node semantics into a single transition relation in order to represent their behavior. 
Assume that the script variable $var$ can be modified by the node \textsf{NodeA} and \textsf{NodeB}, and their node semantics specify that $var$ changes to the same value $val$ in conditions $cond_A$ for \textsf{NodeA} and $cond_B$ for \textsf{NodeB} as follows:
\begin{align*}
var'   & \equiv \text{if } cond_A(in_A, state_A, var) \text{ then } val, \text{ and} \\
var'   & \equiv \text{if } cond_B(in_B, state_B, var) \text{ then } val.
\end{align*}
Then we can integrate those transition relations into the following single relation:
\begin{align*}
var'   & \equiv \text{if } cond_A(in_A, state_A, var) \vee cond_B(in_B, state_B, var) \\
& \quad \text{ then } val.
\end{align*}

\subsection{Scope and Limitations}\label{Subsec:ScopeandLimitations}
\subsubsection{Soundness}\label{Subsec:Soundness}
Strictly speaking, the behavior of our model is not completely equivalent to the actual behavior of the target visual scripts.
This difference may result in false positives or false negatives in verification.
These problems arise due to differences in the time required for signal propagation and abstractions in the modeling of external components.\\

\par\noindent
\textbf{signal propagation}\quad
Even if signal propagation is not time-consuming in the visual script implementation, it takes at least one step of state transition to propagate a signal in our model.
This leads to a case where a particular one of several control flows, which are to be completed simultaneously (or whichever may be finished first), will always be completed first in the model.
As a result, errors that occur only in a particular control flow may be overlooked.

For example, the following two control flows may occur in the script of Fig. \ref{Fig:VisualScriptTypicalBug}:
\begin{itemize}
	\item \textsf{Movie Clip}: \textsf{Finished} $\rightarrow$ \textsf{Set Event Mode}: \textsf{Disable}, and
	\item \textsf{Movie Clip}: \textsf{Skipped} $\rightarrow$ \textsf{If}: \textsf{True} $\rightarrow$ \textsf{Set Event Mode}: \textsf{Disable}.
\end{itemize}
Although both will be completed at the same time in the actual implementation, the former will always be completed faster than the latter in our model.

\par\noindent
\textbf{external components}\quad
Our model does not consider the detailed behavior of external components.
This is because the behavior of external components is often not fully documented and it is difficult to properly model such behavior.
In such cases, nondeterminism can be used to abstractly represent the behavior of external components.

For example, the \textsf{Movie Clip} node in Fig. \ref{Fig:VisualScriptTypicalBug} skips or finishes the playback of the movie depending on the behavior of the external components such as movie player application and game player's input.
We model the behavior as a state transition where \textsf{Playing} changes to one of the states \{\textsf{Playing}, \textsf{Finished}, \textsf{Skipping}\} non-deterministically.
A similar abstraction is used in the definition of node semantics that depend on data flow as shown in \ref{SubSection:ConditionalBranches}.

This abstraction causes the model to include behaviors that do not exist in the actual script, which may result in pseudo-counterexamples.

\subsubsection{Coping with State Explosion}
If we model a huge visual script as it is, its state space grows exponentially, resulting in a state explosion.
Therefore, it is necessary to reduce the size of the model by appropriately partitioning and abstracting the script according to the properties to be checked.
The number of variables in an SMV model have a large impact on the size of the state space to be searched.
In the next section, we will discuss a model optimization to reduce the number of variables in an SMV model.

\subsubsection{Scope of Verification}
By modeling rigorously, many of the above problems related to soundness can be avoided.
However, rigorous modeling increases the model size, which leads to state explosion, and we accept this risk in our verification framework.
In fact, our method currently aims at detecting obvious mis-descriptions of visual scripts as stated in \ref{SubSec:TypicalBugs}, and this risk is not considered to be a practical problem.
We will discuss this issue in Section \ref{Sec:PreliminaryEvaluation}.

\section{State Space Reduction by Optimizing Model}\label{Sec:StateSpaceReduction}
As the size of the node graph increases, the models generated also become larger, making it difficult to perform verification in a realistic amount of time.
To finish the verification process in realistic time, optimization of the model for state space reduction is required.
Since the size of state space increases exponentially with the number of SMV variables, our optimization focuses on how to reduce the number of SMV variables.

\subsection{Remove Nodes without Side Effects}
Observation on the scripts written in our game development shows that most of the nodes have the following characteristics:
\begin{itemize}
	\item Many nodes have multiple inputs and single output.
	\item Many nodes send an output signal after receiving an input signal without side effects such as sending data flow signals or changing values of global variables.
\end{itemize}
We refer such nodes as \emph{NoSE} nodes.
NoSE node can be removed from the node graph by directly reconnecting the ports connected to its input ports with the port connected to its output port.

\begin{figure}[htb]
	\begin{center}
		\includegraphics[width=0.85\linewidth]{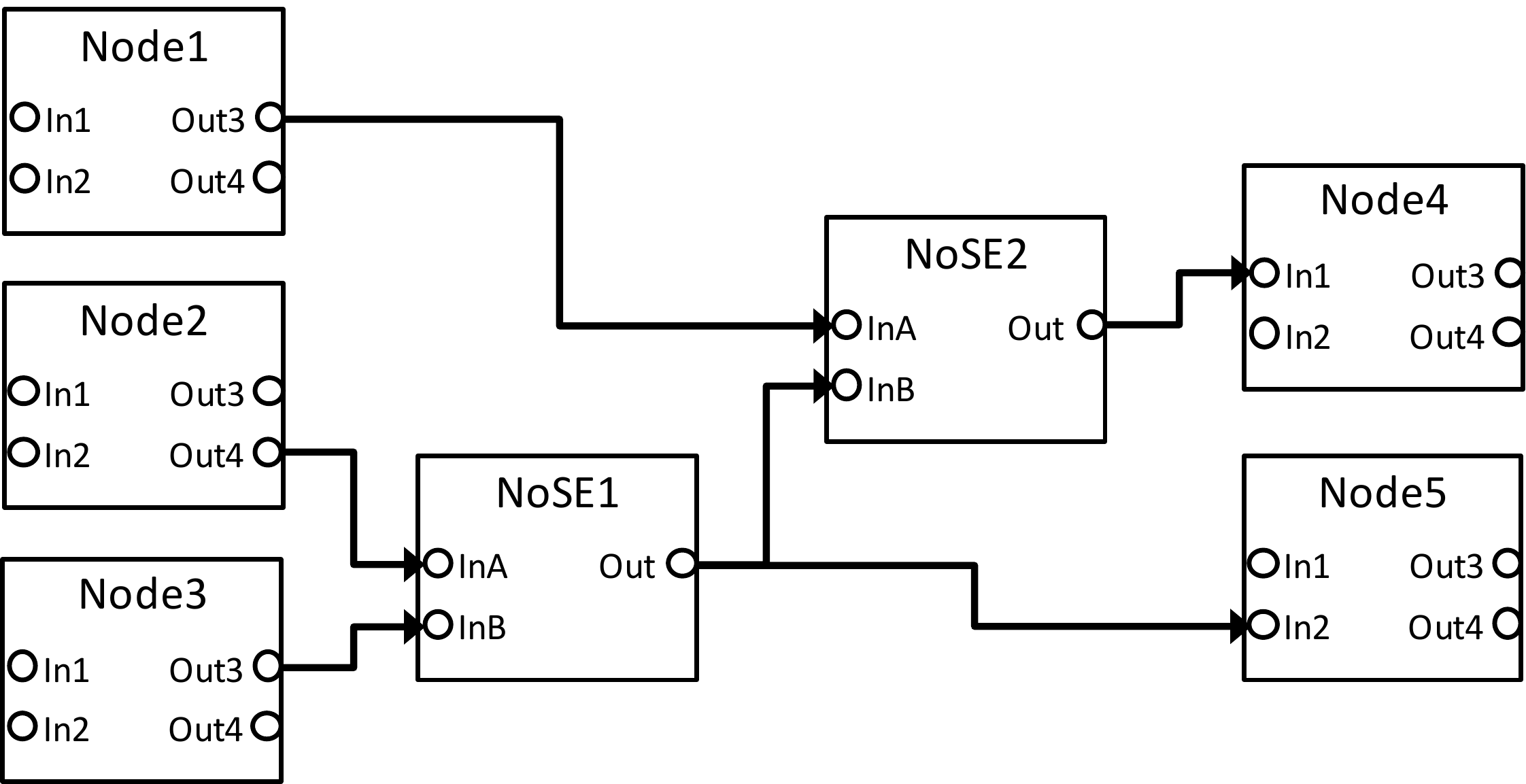}\\
		(a) A graph with NoSE nodes\\
		\vspace{1.0em}
		\includegraphics[width=0.85\linewidth]{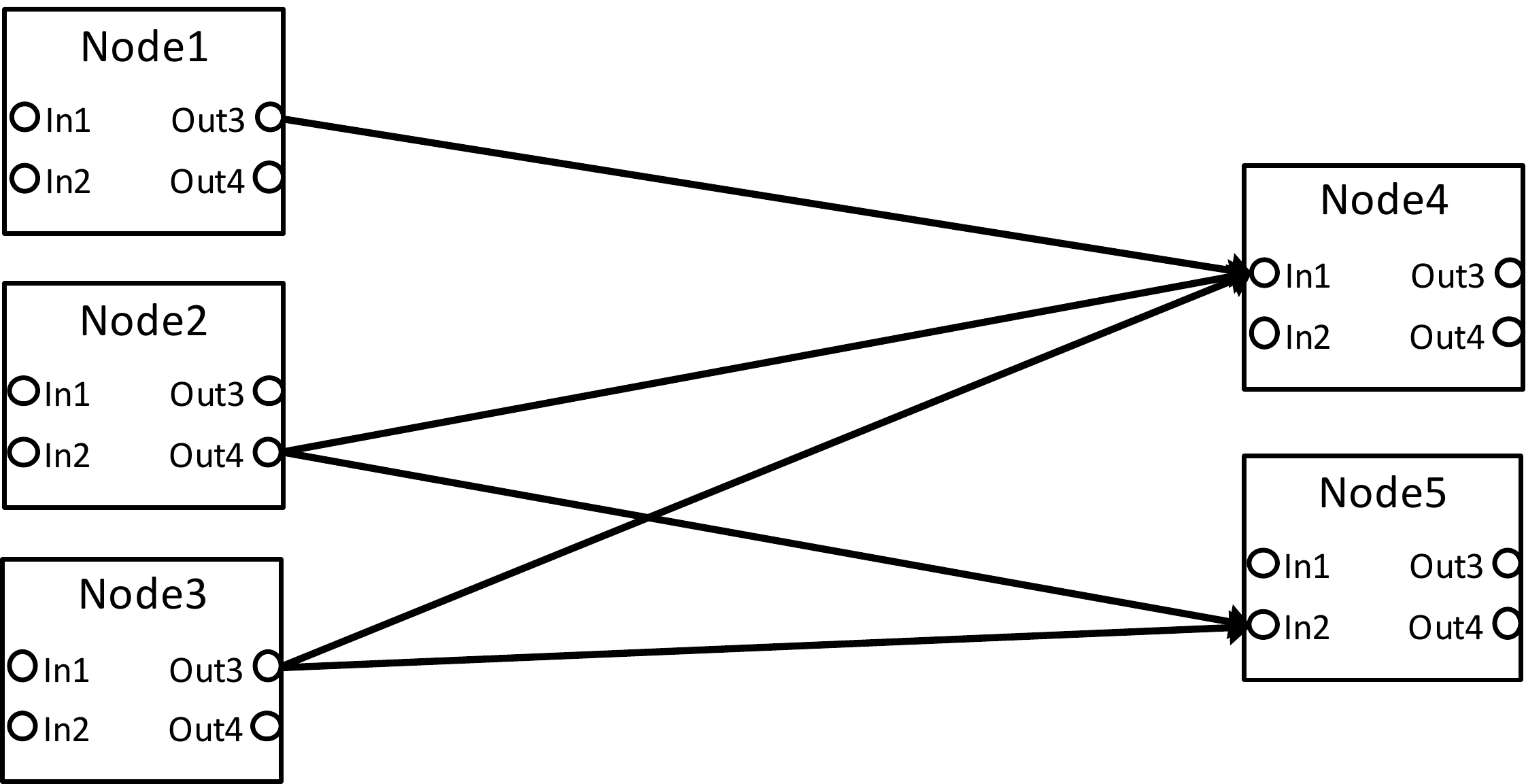}\\
		(b) A graph after removing NoSE nodes
	\end{center}
	\caption{Removing NoSE nodes}\label{Fig:NodeModelOpt}
\end{figure}

For example, Fig. \ref{Fig:NodeModelOpt} (a) shows the node graph which has two NoSE nodes \textsf{NoSE1} and \textsf{NoSE2}.
In this case, we first remove the \textsf{NoSE1} node by reconnecting \textsf{Out4} of \textsf{Node2} and \textsf{Out4} of \textsf{Node3} with \textsf{inB} of \textsf{NoSE2} and \textsf{in2} of \textsf{Node5}.
Then we remove the \textsf{NoSE2} node by \textsf{Out3} of \textsf{Node1} and the two ports connected to its \textsf{inB} port with \textsf{in1} of \textsf{Node4}.
Thus we can obtain the node graph shown in Fig.\ref{Fig:NodeModelOpt} (b).

Since the number of SMV variables is proportional to the number of (ports of) nodes, the number of SMV variables can be greatly reduced by removing those many harmless nodes.

Strictly speaking, applying this optimization can slightly change the behavior of the model for the same reason as stated in \ref{Subsec:ScopeandLimitations}.
We also accept this risk by the same reason.

\subsection{Encode State Transition to Output Variable}
We explained how our method handle state transition of nodes in \ref{SubSection:StateTransition}.
However, this increases the number of SMV variables and makes a negatively impact to the verification performance.
We will explain the optimization that embeds the state transition of the node to the output variable,
and show that it can reduce the number of SMV variables.

As stated in \ref{SubSec:NodeSemantics}, the node semantics is generally expressed as the following formulas:
\begin{align}
\label{formula:out0} out'   & \equiv f_{out}(in, state), \\
\label{formila:state0} state' & \equiv f_{state}(in, state).
\end{align}
The value of the output signal only depends on its internal state in the node with state transition.
Furthermore, if we were strictly represent the behavior of the node, the output signal should change immediately according to the change of internal state.
We can thus obtain the following relation instead of \ref{formula:out0}:
\begin{equation}
\label{formula:out1} out \equiv f_{out}(state).
\end{equation}

We focus on the fact that the internal state is associated with the value of the output signal shown in \ref{formula:out1}.
Therefore, by representing the internal state using the value of the output signal, the state variable can be removed from the equation.
The basic idea of this optimization is the following derivation:
\begin{enumerate}
	\item Apply the inverse function of $f_{out}$ to both sides of the equation \ref{formula:out1}.
	\begin{equation}
	\label{formila:state1} state \equiv f_{out}^{-1}(out)
	\end{equation}
	\item Substitute \ref{formila:state1} for \ref{formila:state0}.
	\begin{equation}
	\label{formula:state2} f_{out}^{-1}(out') \equiv f_{state}(in, f_{out}^{-1}(out))
	\end{equation}
	\item Apply $f_{out}$ to both sides of the equation \ref{formula:state2}.
	\begin{equation}
	\label{formula:out2} out' \equiv f_{out}(f_{state}(in, f_{out}^{-1}(out)))
	\end{equation}
\end{enumerate}
As shown in \ref{formula:out2}, we can replace state variables with output variables with this derivation.
We thus can reduce the number of SMV variables.

Since $f_{out}$ is a surjective function, but not an injective function, we can not define the inverse function of $f_{out}$.
Then, we define a bijective function $g_{out}$ by extending $f_{out}$.
When an output signal is associated to multiple internal states, we duplicate the output signal in order to associate them with single internal state.
The duplicated signal of $out$ associated to the state $state$ is referred as $out_{state}$.
We define the bijective function $g_{out}$ as follows:
\[
g_{out}(state) =
\begin{cases}
out & \text{if for all internal states } t,\\
& state \neq t \rightarrow f_{out}(state) \neq f_{out}(t) \\
out_{state}     & \text{otherwise}
\end{cases}
\]
where $out = f_{out}(state)$.
$out_{state}$ is a newly defined value that is different from $out$, but is treated in the same way as $out$.

Using the bijective function $g_{out}$, we can redefine the formula \ref{formula:out2} as follows:
\begin{equation}
\label{formula:out3} out' \equiv g_{out}(f_{state}(in, g_{out}^{-1}(out)))
\end{equation}
We refer this function as $h_{out}(in, out)$.

For example, the transition relation $f_{out}(state)$ and $f_{state}(in, state)$ of the \textsf{Movie Clip} node is defined as follows:
\begin{align*}
f_{out}(state) & \equiv \text{if } state = \textsf{Finished} \text{ then } \textsf{Finished}\\
& \quad \text{else if } state = \textsf{Skipped} \text{ then } \textsf{Skipped}\\
& \quad \text{else } none\\
f_{state}(in, state) & \equiv \text{if } in = \textsf{Start} \text{ then } \textsf{Playing}\\
& \quad \text{else if } state = \textsf{Playing}\\ 
& \quad \quad \text{then } \{\textsf{Playing}, \textsf{Finished}, \textsf{Skipped}\}\\
& \quad \text{else } \textsf{Stopped}
\end{align*}

We can obtain $g_{out}$ and $g_{out}^{-1}$ from $f_{out}$ as follows:
\[
g_{out}(state) \equiv
\begin{cases}
none_{\textsf{Stopped}} & \text{if } state = \textsf{Stopped} \\
none_{\textsf{Playing}} & \text{if } state = \textsf{Playing} \\
\textsf{Finished}       & \text{if } state = \textsf{Finished} \\
\textsf{Skipped}        & \text{if } state = \textsf{Skipped}
\end{cases}
\]
\[
g_{out}^{-1}(out) \equiv 
\begin{cases}
\textsf{Stopped}  & \text{if } out = none_{\textsf{Stopped}} \\
\textsf{Playing}  & \text{if } out = none_{\textsf{Playing}} \\
\textsf{Finished} & \text{if } out = \textsf{Finished} \\
\textsf{Skipped}  & \text{if } out = \textsf{Skipped}
\end{cases}
\]

Finally we can obtain the optimized transition relation $h_{out}(in, out)$ as follows:
\begin{align*}
h_{out}(in, out) & \equiv \text{if } in = \textsf{Start} \text{ then } none_{\textsf{Playing}}\\
& \quad \text{else if } out = none_{\textsf{Playing}} \\ 
& \quad \quad \text{then } \{none_{\textsf{Playing}}, \textsf{Finished}, \textsf{Skipped}\}\\
& \quad \text{else } none_{\textsf{Stopped}}
\end{align*}

In addition, we need to modify the set of initial values of the output signal.
We refer the set as $I_{out}$.
Since the internal state of the node is identified by the value of its output signal, $I_{out}$ is defined from the set of initial values of internal states of the node $I_{state}$ and $g_{out}$ as follows:
\begin{equation}
\label{formula:init} I_{out} \equiv \{g_{out}(s) \mid s \in I_{state}\}.
\end{equation}

Fig. \ref{Fig:OptMovieClipNodeModel} shows the model that we applied this optimization to the \textsf{Movie Clip} node.
As you can see, we can remove the state variable \texttt{MovieClip3State}.

\begin{figure}[htb]
\begin{lstlisting}[]
VAR
  MovieClip3In : {none, Start};
  MovieClip3Out : {none_Stopped, none_Playing, Finished, Skipped};
ASSIGN
  init(MovieClip3Out) := none_Stopped;
  next(MovieClip3Out) := case
    MovieClip3In = Start : none_Playing;
    MovieClip3Out = none_Playing :  {none_Playing, Finished, Skipped};;
    TRUE : none_Stopped;
  esac;
\end{lstlisting}
\caption{Optimized \textsf{Movie Clip} node model}\label{Fig:OptMovieClipNodeModel}
\end{figure}

\section{Preliminary Evaluation}\label{Sec:PreliminaryEvaluation}

\begin{table*}[t]
	\caption{Preliminary evaluation of our method}\label{Table:EvalMethod}
	\begin{center}
		\def\arraystretch{1.2}
		
		\begin{tabular}{|l|c|c|c|c|c|c|c|c|c|c|} \hline
			& & \multicolumn{3}{|c|}{without  optimization} & \multicolumn{5}{|c|}{with optimization} & \\ \hline
			\# & \# of nodes & \# of vars &  reachable states & eval. time[s] & \# of vars &  \multicolumn{2}{|c|}{reachable states} & \multicolumn{2}{|c|}{eval. time[s]} &  detected? \\ \hline \hline
			\#1 & 156 & 356 &  $2^{23.0783}$ & 192.786 & 180 &  $2^{16.6621}$ & $\downarrow$ 98.8 \% & 3.603& $\downarrow$ 98.1 \% & no \\ \hline
			\#2 & 94  & 214 &  $2^{13.2429}$ & 3.330   & 121 &  $2^{12.729}$  & $\downarrow$ 30.0 \% & 0.994& $\downarrow$ 70.2 \% & no \\ \hline
			\#3 & 37  & 84  &  $2^{4.32193}$ & 0.056   & 38  &  $2^{3.80735}$ & $\downarrow$ 30.0 \% & 0.037& $\downarrow$ 33.9 \% & no \\ \hline
			\#4 & 49  & 119 &  $2^{9.06609}$ & 0.111   & 59  &  $2^{8.86419}$ & $\downarrow$ 13.1 \% & 0.060& $\downarrow$ 45.9 \% & no \\ \hline
			\#5 & 177 & 414 &  $2^{11.412}$  & 36.675  & 219 &  $2^{10.9694}$ & $\downarrow$ 26.4 \% & 10.169& $\downarrow$ 72.3 \% & ok \\ \hline
			\#6 & 73  & 162 &  $2^{8.01681}$ & 0.173   & 82  &  $2^{7.41785}$ & $\downarrow$ 34.0 \% & 0.088& $\downarrow$ 49.1 \% & no \\ \hline
			\#7 & 162 & 408 &  $2^{17.9397}$ & 98.102  & 199 &  $2^{17.3921}$ & $\downarrow$ 31.6 \% & 17.103& $\downarrow$ 82.6 \% & ok \\ \hline
			\#8 & 430 & 980 &  N/A & N/A & 475 & N/A & N/A & N/A & N/A & N/A \\ \hline
		\end{tabular}
		
		Env.: Intel(R) Core(TM) i7-3770 CPU @ 3.40GHz / 32GB / Windows 7 (64 bit) / NuSMV 2.6.0
	\end{center}
\end{table*}

For a preliminary evaluation, we implemented a prototype and applied it to the visual scripts that are used in the production for \FFXV and perform flag management.
Seven scripts (\#1 - \#7) were randomly selected, and one very large script (\#8) was arbitrarily selected to clarify the limit of script size to which our method can be applied.
All of these scripts handle the \textsf{event mode} flag, and we checked the property that if the flag is true, it must eventually become false.
The CTL formula annotated in the SMV model is the same as the one shown in \ref{Sec:Specification}. 
We also applied the proposed optimization to reduce the size of the SMV models.

\subsection{Node Semantics}\label{SubSec:NodeSemanticsForEvaluation}
We prepared an encoding by SMV language for each node in the scripts.
As stated in section \ref{SubSec:NodeSemantics}, we can straightforwardly prepare an encoding for nodes with simplified semantics.
The eight scripts shown in Table \ref{Table:EvalMethod} have 1,178 nodes in total.
There are 164 kinds of those nodes, which are classified as follows:
\begin{enumerate}
	\item node with single output: 98 kinds of nodes.
	\item node with conditional branches : 7 kinds of nodes.
	\item node with state-transition: 14 kinds of nodes.
	\item node with conditional branches and state-transition: 12 kinds of nodes.
	\item entry point node: 3 kinds of nodes.
	\item node with custom semantics: 30 kinds of nodes.
\end{enumerate}
Nodes classified from 1) to 5) can be modeled by using a common node semantics for each classification.
Additionally we have prepared custom semantics for 30 kinds of nodes in 6).
In the end, we can convert 1,178 nodes to SMV models by manually preparing only 30 types of node semantics (furthermore, these node semantics may be reused).
This result demonstrates that our translation method has enough availability in practical use.

\subsection{Verification Results}\label{SubSec:Result}
Table \ref{Table:EvalMethod} shows the results of the evaluation.
The column descriptions are the following:
\begin{itemize}
	\item \# of nodes: The number of visual script nodes in the target script.
	\item \# of vars: The number of SMV variables in the generated SMV model.
	\item reachable states: The number of reachable states reported by NuSMV.
	\item eval. time: Execution time of NuSMV for the model.
	We tried 5 times for each script and adopted a median value of those trial.
	\item detected?: Whether NuSMV detected a problem in the script or not.
\end{itemize}

We can detect the bugs on two scripts \#5 and \#7 and can obtain the counterexamples.
As you can see, we can significantly reduce the verification time by applying the optimization.
However, the verification of \#8 was not completed within the 3 hours set as the timeout.

\subsection{Discussion}
\subsubsection{soundness}
We found counterexamples on two scripts \#5 and \#7 during the experiments.
We confirmed with the game designers that those counterexamples are not false positive and indicate real bugs in the scripts.
\footnote{According to the game designers, those scripts are used only in the trial version, so they will not fix the bugs though.}
This result demonstrates that our method can detect the specific types of bugs that we are focusing on.

\subsubsection{optimization effectiveness}
By applying the optimization, we can considerably reduce the state space to be checked and the verification cost.
Particularly we can decrease the verification time by 98.1 \% in the case of \#1.
Besides, the effect of optimization in the two cases where the script bugs were detected is greater than in the other scripts.
This result suggests that the effect of optimization is greater in terms of bug detection.

\subsubsection{state explosion}
This result showed that even with optimization, it is difficult to handle very large scripts in our framework.
By partitioning the node-graph according to the structure of the control flow and applying further abstraction, we expect that it is possible to reduce the state space and achieve verification in a realistic amount of time even for such a large script.
Improving our algorithm to handle those large scripts is future work.

\section{Summary and Future Work}\label{Sec:Summary}

We described an automatic verification method for node-based visual script notation for efficient game production.
Our method automatically converts visual script implementation with node semantics to the input model for NuSMV.
Preliminary evaluations confirmed that for most of the visual scripts used in the production of \FFXV, we could detect the specific types of bugs we were focusing on in a realistic amount of time.

However, it appears that there are some very large scripts used in the production for \FFXV, that our method cannot handle.
Thus a next step for improving our work would be applying compositional verification technique \cite{Berezin1998}.
If the node graph can be partitioned into multiple subgraphs, and each subgraph can be verified separately, then the verification cost can be reduced and the verification can be completed in a realistic time.
In addition, compositional verification also may allow for the verification of game logic written as multiple scripts.
Currently, we can verify only scripts written as a single node graph.
If multiple scripts can be verified together, the control flow between scripts can be tracked, resulting in a reduction of false positives and false negatives.
Another next step would be the automated generation of node semantics from the script implementation.
Currently, we need to write node semantics manually.
If we can extract node semantics from its implementation, we can increase the range of automation of our method.

\section*{Acknowledgment}
We wish to thank the collaborative researchers for helpful discussions.
We also wish to thank \FFXVDevTeam \ for supporting our research.
UNREAL ENGINE is a trademark or registered trademark of Epic Games, Inc.
Windows is a trademark or registered trademark of Microsoft Corporation.
All other trademarks are the property of their respective owners.

\bibliographystyle{IEEEtran}
\bibliography{myref_arXiv}

\end{document}